\documentclass[letterpaper]{article} 
\usepackage{aaai25}  
\usepackage{times}  
\usepackage{helvet}  
\usepackage{courier}  
\usepackage[hyphens]{url}  
\usepackage{graphicx} 
\usepackage{natbib}  
\usepackage{caption} 
\usepackage{algorithm}
\usepackage{algorithmic}

\usepackage{subcaption}
\usepackage{booktabs} 
\usepackage{xcolor} 
\usepackage{amsmath} 

\usepackage{newfloat}
\usepackage{listings}
\DeclareCaptionStyle{ruled}{labelfont=normalfont,labelsep=colon,strut=off} 
\floatstyle{ruled}
\newfloat{listing}{tb}{lst}{}
\floatname{listing}{Listing}
\title{Can Platform Design Encourage Curiosity? \\ Evidence from an Independent Social Media Experiment}
\author {
    Marie Neubrander\equalcontrib\textsuperscript{\rm 1},
    Markus Reiter-Haas\equalcontrib\textsuperscript{\rm 2},
    Ben Rochford\textsuperscript{\rm 3},
    Max Allamong\textsuperscript{\rm 4},
    Christopher Bail\textsuperscript{\rm 3}, \\
    Sunshine Hillygus\textsuperscript{\rm 5},
    Alexander Volfovsky\textsuperscript{\rm 1}
}
\affiliations {
    \textsuperscript{\rm 1} Department of Statistical Science, Duke University\\
    \textsuperscript{\rm 2} Institute of Human-Centered Computing, Graz University of Technology\\
    \textsuperscript{\rm 3} Department of Sociology, Duke University\\
    \textsuperscript{\rm 4} Department of Political Science, University of Houston\\
    \textsuperscript{\rm 5} Department of Political Science, Duke University\\
    \{marie.neubrander, ben.rochford, christopher.bail, hillygus, alexander.volfovsky\} @duke.edu, reiter-haas@tugraz.at, mballamong@uh.edu
}

\usepackage{bibentry}

\begin{document}

\maketitle

\begin{abstract}
Social media platforms are often criticized for fostering antisocial behavior rather than prosocial behavior. Yet, testing interventions to encourage prosocial dispositions, such as open-mindedness, has been hindered by researchers' limited ability to manipulate platform features and isolate causal effects in commercial environments. We address this challenge through a randomized controlled trial with 2,282 U.S. adults conducted on a new research platform we developed that uses AI bots to replicate live social media dynamics while enabling controlled experimentation. Participants engaged in 15-minute discussions about energy and climate topics, with treatment groups exposed to curiosity priming either through modified on-platform social norms, interface affordances, or both. Results demonstrate that curiosity priming significantly increased question-asking behavior and textual measures of curiosity in user posts, while also reducing toxicity. Although interventions decreased generic engagement behaviors like liking and commenting, they had no significant negative impact on reported app enjoyment or time spent writing posts and replies. Leveraging experimental control over platform features, our findings suggest that platform designs prioritizing curiosity can promote prosocial behaviors among users without compromising user experience.
\end{abstract}

%

\section{Introduction}
Social media platforms have become central venues for public discourse, yet mounting evidence suggests they often amplify antisocial behaviors instead of facilitating productive exchange. Research has documented, for example, how platform dynamics can exacerbate affective polarization \cite{bail2018exposure, finkel_political_2020}, encourage performative outrage \cite{van_bavel_social_2024,brady_how_2021}, and foster echo chambers that insulate users from diverse perspectives \cite{pariser_filter_2012, sunstein_republic_2017, gonzalez-bailon_asymmetric_2023, boulianne2024social}. In response, scholars across the computer and social sciences have increasingly focused on interventions to mitigate these harms and, more ambitiously, to cultivate prosocial behaviors---actions intended to benefit others or society as a whole \cite{dorr_research_2025, celadin_promoting_2024, combs_reducing_2023, taylor_accountability_2019, prot_long-term_2014, rajadesingan_guessync_2023, gruning_framework_2024, gruning_independently_2024, van_loon_designing_2025, larooij_can_2025}. Prosocial behaviors in online contexts include expressing empathy, engaging constructively across disagreement, and demonstrating openness to alternative viewpoints \cite{nook_prosocial_2016, dorr_research_2025}. Such behaviors are thought to be critical for maintaining democratic discourse and reducing the affective divides that characterize contemporary political life \cite{iyengar_affect_2012, iyengar_origins_2019, finkel_political_2020, hartman_interventions_2022, bail_breaking_2021}.

One promising mechanism for fostering prosociality is the promotion of curiosity in social media discourse. When people attempt to learn from others rather than defend their own views, interactions should become less combative and hostile, encouraging more productive discourse \cite{dorr_research_2025, leary_cognitive_2017, yu_measuring_2025}. Curiosity, and particularly ``epistemic curiosity,'' the desire to acquire new knowledge, has been shown to promote intellectual humility, the recognition of one’s own fallibility and openness to revising beliefs \cite{leary_cognitive_2017}. Individuals high in intellectual humility are more likely to befriend ideological opponents and less likely to form echo chambers \cite{stanley_intellectual_2020}. Question-asking, a behavioral manifestation of curiosity, signals conversational receptiveness and facilitates perspective-taking \cite{huang_it_2017, chen_tell_2010}. Curiosity can be understood as a motivational foundation for open-minded thinking, a willingness to consider alternative perspectives and change one’s mind in light of new evidence. Actively open-minded thinking has been shown to reduce polarization and improve reasoning quality \cite{baron_actively_2019}. While numerous prosocial outcomes merit investigation, curiosity stands out for its downstream benefits for discourse quality.

Despite this theoretical promise, it remains unclear if and how social media platforms could encourage curiosity. For example, could a platform nudge curiosity by replacing a neutral post prompt of ``What's on your mind?'' with one that asks ``What are you curious about?'' Unfortunately, conducting such an experiment on commercial platforms is difficult. Researchers typically lack access to modify affordances or randomize users to different interface designs \cite{orben_fight_2025, orben_fixing_2025, freelon_post-api_2024}. Even when platform partnerships exist, they can impose constraints on experimental design and unforeseen complexities can limit generalizability \cite{guess_how_2023, bagchi_social_2024}. Both social norms and affordances shape user behavior, yet systematically testing their effects requires control that commercial platforms rarely grant. These challenges have intensified as platforms have restricted API access and increased data costs, with services like Twitter/X and Reddit now charging tens of thousands of dollars for full research access \cite{freelon_post-api_2024}. A growing methodological response involves independent research platforms---controlled environments that replicate social media experiences while enabling careful control of platform design factors and randomized experimentation \cite{gruning_independently_2024, allamong_causal_2025, van_loon_designing_2025, taylor_accountability_2019, epstein_how_2022, aghajari_investigating_2024}.

In this context, we contribute a three-arm randomized controlled trial with 2,282 U.S. adults. We leverage a new research platform augmented with AI bots to emulate realistic social media dynamics. Participants discussed energy and climate topics for 15 minutes under one of three conditions: a control resembling standard social media, a treatment priming curiosity through modified social normative cues in onboarding instructions, and a treatment combining these normative cues with altered affordances including revised post prompts and a ``Thought-provoking'' reaction button. We pose five distinct research questions: 
\begin{description}
    \item[RQ1:] Does curiosity priming increase question-asking behavior and textual measures of curiosity?
    \item[RQ2:] Does curiosity priming reduce toxicity in user contributions?
    \item[RQ3:] Does curiosity priming affect overall engagement volume?
    \item[RQ4:] Does curiosity priming impact users' enjoyment of the platform?
    \item[RQ5:] Does curiosity priming impact users' measured intellectual humility?
\end{description}

Our results show that curiosity priming significantly increases question-asking and curiosity scores while reducing toxicity, with no detrimental effects on engagement or user satisfaction. Interestingly, we find no significant effects on self-reported intellectual humility, suggesting that behavioral shifts in curiosity may not immediately translate to broader attitudinal changes. These findings have practical implications for platform design and moderation strategies, and demonstrate how independent research platforms can support causal inference in social media research.

\section{Related Work}

While there is a rich body of work that explores the relationship between social media and societal harms, the current study builds upon three related literatures in motivating our design and expectations.

\subsection{Prosocial Interventions on Social Media}
A growing body of research examines interventions designed to promote prosocial behavior on social media platforms. \citeauthor{gruning_independently_2024} provide a framework for digital interventions that promote prosocial outcomes, underscoring the need for experimental research. Recent work has explored diverse intervention strategies. \citeauthor{celadin_promoting_2024} conducted a tournament comparing seven nudge-based interventions to promote civil discourse, finding that accuracy prompts and social norms cues showed promise. \citeauthor{pretus_misleading_2024} demonstrated that identity-based interventions can reduce partisan misinformation sharing by highlighting in-group members who rated content as misleading. Other studies have targeted specific antisocial behaviors, with \citeauthor{taylor_accountability_2019} designing accountability features to encourage bystander intervention against cyberbullying. Despite this progress, most interventions have focused on reducing specific harms such as misinformation, incivility, and toxicity. Interventions designed to cultivate prosocial dispositions have the potential to more durably improve online discourse by moving beyond simple prevention and mitigation \cite{herzog_boosting_2025, dorr_research_2025}.

\subsection{Curiosity and Intellectual Humility}
Among potential prosocial constructs, intellectual humility and curiosity have emerged as particularly promising targets for intervention. Research shows that intellectual humility, the acknowledgment of one's limitations and willingness to update one's views, is associated with decreased polarization and stronger cross-partisan dialogue---a central concern in the social media literature \cite{stanley_intellectual_2020, baron_actively_2019}. The desire to acquire new knowledge appears to cultivate this humility by driving individuals to pursue diverse information sources and genuinely engage with competing viewpoints \cite{leary_cognitive_2017}. The act of posing questions demonstrates openness to learning and creates opportunities for understanding others' reasoning \cite{huang_it_2017, chen_tell_2010}.

Empirical evidence on curiosity-promoting interventions remains limited. Existing work has largely examined structural features that might indirectly foster curiosity. For instance, \citealt{rajadesingan_walking_2021} identified design strategies for facilitating cross-partisan discussions, while \citealt{rajadesingan_guessync_2023} developed a casual game to reduce affective polarization through perspective-taking exercises. \citeauthor{van_loon_designing_2025} tested how different status affordances affect political dialogue quality. However, none of these studies directly manipulated curiosity itself or systematically measured its behavioral manifestations like question-asking. This gap is critical given curiosity's theoretical promise: if platform designs can successfully prime curiosity, they might simultaneously reduce toxicity while increasing the quality of discourse through more questioning and open-minded engagement \cite{huang_it_2017, chen_tell_2010, leary_cognitive_2017}.

\subsection{Independent Platforms for Social Media Research}
Creating tests of platform-level interventions is methodologically challenging. Commercial platforms rarely grant researchers the access needed to manipulate affordances or randomize users to different interface designs \cite{orben_fight_2025, orben_fixing_2025, freelon_post-api_2024}. Even successful industry partnerships impose constraints on experimental design and involve unknowns which can raise questions about generalizability \cite{gonzalez-bailon_asymmetric_2023, guess_how_2023, bagchi_social_2024}. These challenges have intensified as platforms restrict API access and increase data costs \cite{freelon_post-api_2024}.

Independent research platforms offer a methodological solution by replicating social media experiences in controlled environments while enabling randomized experimentation \cite{gruning_independently_2024, allamong_causal_2025, van_loon_designing_2025, taylor_accountability_2019, epstein_how_2022, aghajari_investigating_2024}. One approach uses fully synthetic environments where generative AI agents simulating social dynamics, as demonstrated by \citeauthor{larooij_can_2025}. A separate line of research has developed platforms where real participants interact in controlled social media environments. \citeauthor{combs_reducing_2023} tested whether such a platform could reduce polarization through cross-partisan conversation. More recently, this approach has been extended by integrating AI bots alongside real users. Forthcoming research \cite{van_loon_designing_2025} has examined how status-related affordances affect political dialogue on a platform populated in part by AI bots, and orchestrated bot behavior to causally examine the impacts of incivility on social media \cite{allamong_causal_2025}.

Building upon this line of work, we leverage a research platform we developed which integrates AI bots with real user participation. We conduct a randomized controlled trial that manipulates curiosity through both normative cues and interface affordances within a controlled yet realistic social media environment. We provide the first causal evidence on whether platform design can prime curiosity, and examine how such interventions impact behavior to provide an empirical examination of theoretical claims about curiosity's prosocial benefits.

\section{Methods}

\subsection{Platform}

The study was conducted using a new research platform we developed \footnote{Link to the platform webpage removed for peer review.} to provide participants with a believable simulation of a live social media network within a controlled environment. Participants are recruited to test a new social media platform called ``Spark Social.'' While participants believe they are interacting with other users, the accounts in their feed are actually synthetic profiles powered by AI bots using OpenAI’s `gpt-4-turbo-preview' Large Language Model \cite{achiam2023gpt}. This enables us to manipulate not only platform features, but also the attributes of the synthetic profiles (such as username, political affiliation, and or occupation). Participants experience these features of the bot profiles through the style and content of their generated posts and comments. 

\begin{figure*}[t] 
    \centering
    \includegraphics[width=\textwidth]
    {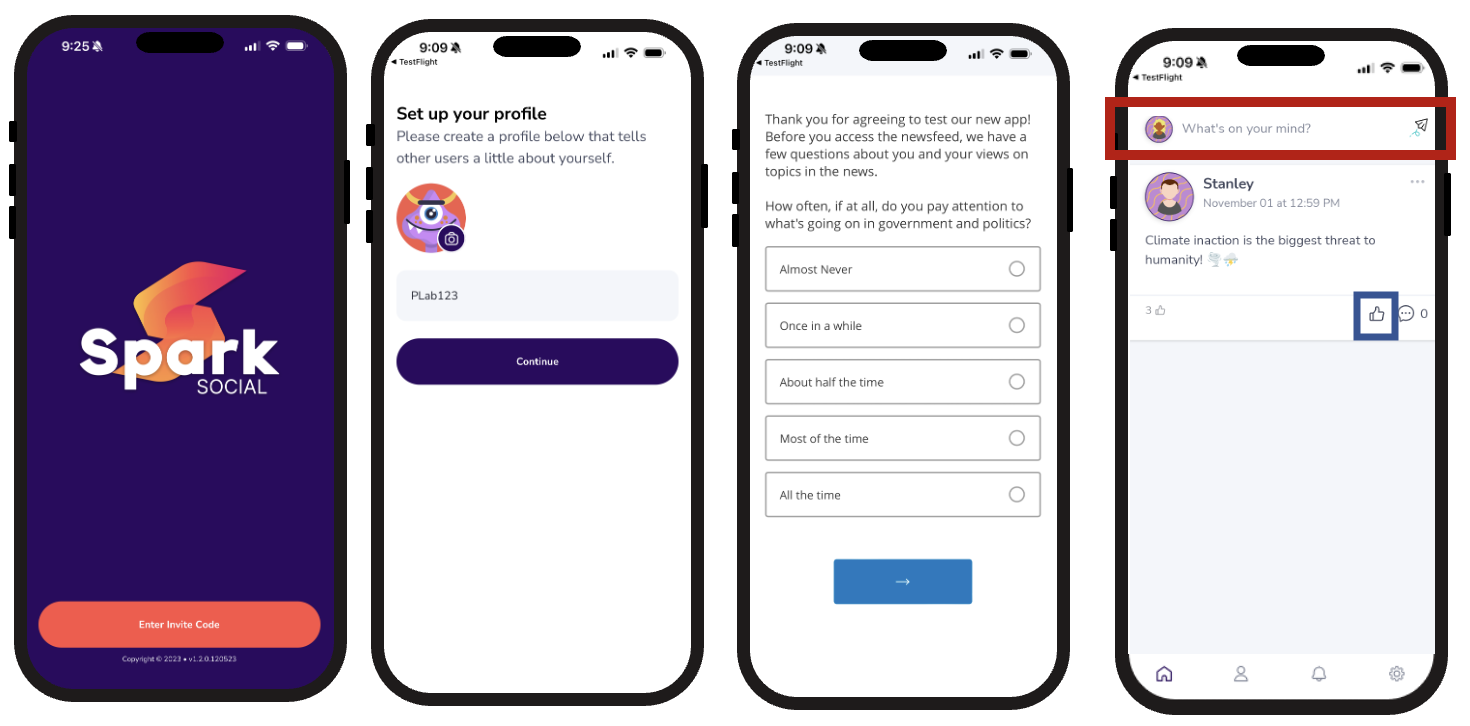}
    \caption{Screenshots from the mobile application used by participants. Figure shows the landing page (far left), the profile creation page (middle left), the embedded survey (middle right), and the newsfeed (far right). Red box highlights the text input prompt; blue box highlights the ``Like'' button. }
    \label{fig:screenshots}
\end{figure*}

From a participant's perspective, the experience is designed to feel similar to existing social media platforms. To access the study, participants download the ``Spark Social'' application from the Apple App Store or Google Play Store. Upon joining the application, they first take an embedded pre-treatment survey and are then met with a newsfeed of posts from existing accounts. Participants had the ability to read, post, comment, and react (e.g., liking a post). The environment is most similar to platforms where individuals engage with networks of strangers or loose connections, such as Reddit, Threads, or X. Figure \ref{fig:screenshots} demonstrates the mobile application's user interface. Once they receive an invite code through Prolific, participants download the app where they see the landing  page, set up a profile, complete an embedded pre-treatment survey, then spend a predefined amount of time (e.g, 15 minutes) engaging in their newsfeed. Once the time is up, they also completed an embedded post-treatment survey. Participants complete the survey anonymously.

\subsection{Study Design}

In Fall 2024, we recruited 2,282 U.S. adults using the online non-probability panel, Prolific to test a new social media platform. As preregistered, participants who failed attention checks were excluded from our analytic sample (refer to Appendix \ref{app:prereg} for more details).
Users were recruited to test a new social media environment (full details of the recruitment procedure are in Appendix~\ref{app:recruitment}) and took surveys before and after treatment. The study had a total expected time of 30 minutes and participants were compensated with \$7.50 (with the potential for a \$1 bonus), totaling \$25,917.34 in participant compensation across the study (including a small pilot study). Participants were randomly assigned to one of three conditions (instructions and screenshot are in Appendix~\ref{app:arms}): 

\begin{itemize}
    \item \textit{\textbf{Control}} (\(n=771\)): A control condition resembling a mainstream text-based social media platform. Onboarding instructions welcomed users to the conversation, the post input prompt (Figure \ref{fig:screenshots}, red box) asked ``What's on your mind?'' with a paper plane animation, and the reaction mechanism was a standard ``Like'' button (Figure \ref{fig:screenshots}, blue box).

    \item \textit{\textbf{Treatment 1 (T1)}} (\(n=733\)): A \textit{curiosity norm} condition in which users are primed for curiosity. The onboarding instructions introduced users to an app named \textit{Curio}, with a custom question-mark-inspired logo, and instructed users to ``spark curiosity'' and ask questions. The platform interface remained identical to the control setting: the post input prompt asked ``What's on your mind?'', and the reaction mechanism was a standard ``Like'' button.

    \item \textit{\textbf{Treatment 2 (T2)}}  (\(n=763\)): A \textit{curiosity norm and affordances} condition. Users are primed for curiosity in the same manner as in T1. To induce a stronger treatment, interface changes also made. The post input prompt was changed to ``What are you curious about?'' and replaced the default paper plane animation with a question-asking speech bubble animation, while also replacing the ``Like'' button with a ``Thought-provoking'' (lightbulb) button.
\end{itemize}

In each arm, users spent 15 minutes discussing topics related to energy, climate, and the environment with six AI bots, each taking on a specific occupation, U.S. political party, knowledge level (e.g., well-informed), and stance regarding content positivity (e.g., optimistic).  We selected environmental topics because they were prominent in public discourse at the time of recruitment but were not the primary focus of the concurrent election cycle. Table \ref{tab:my_label} in the Appendix provides complete summaries of the six AI bots.

Across all treatment arms, the bots were created with the same prompted instructions and characteristics. Bots were specifically prompted to ask questions.
Additionally, across all treatment arms, each user's feed began with the same set of six pre-populated posts and three pre-populated comments, regardless of treatment arm. Once users began interacting with the platform, feeds and engagement patterns diverged, allowing us to analyze causal impacts of curiosity priming.

\subsubsection{Participant Demographic Data}

Of the 2,282 study participants, the majority were women (58\%) and skewed towards younger and middle aged adults, with over half (55\%) between the ages of 25-44. Nearly half of participants held a bachelor's degree or higher (48\%); another quarter (26\%) have completed at least some college. Politically, the sample was fairly balanced between Democrats (37\%), Republicans (34\%), and independents (28\%). Four participants had missing  values for demographic information. The full breakdown of age, education, gender, political party, and race of the remaining participants is shown in Figure \ref{fig:demographics} in the Appendix.

\subsection{Measured Outcomes and Statistical Analysis}
The research platform used for the study enables the collection and analysis of a wide range of behavioral and attitudinal data. We monitored in-app activity by capturing metrics such as writing time and engagement counts alongside the raw textual content of user posts and comments. Participants additionally completed pre- and post-study surveys to measure any changes in attitudes and to assess the overall app experience. Our analysis focuses on the five research questions detailed in this section below.

\textit{\textbf{RQ1: Curiosity.}} \textit{Does curiosity priming increase question-asking behavior and textual measures of curiosity?} To answer this question, we examine two text-based metrics. The first is question mark usage: we model the presence of at least one question mark in a user's contribution as a binary outcome under the assumption that this serves as a proxy for a participant engaging in question-asking. While the question mark is a simple heuristic likely to miss some question-asking behavior, it serves as an interpretable baseline measurement \cite{huang_it_2017}.

Throughout our analysis, we examine posts and comments separately as these contributions generally take on different communicative roles: initiating versus responding to discussion. To avoid redundancy, this section defines models in terms of user ``contributions'' with the understanding that each model is fit twice independently: once for posts and once for comments.  

Because participants in our study could contribute multiple posts and comments throughout their 15-minute session, the resulting data is nested within individuals. This violates the independence assumption required for standard regression. To account for this, we used Generalized Linear Mixed-Effects Models (GLMMs). GLMMs extend the generalized linear regression framework (e.g., logistic regression) by incorporating random effects. These allow us to estimate the treatment effect while accounting for the baseline variance in individual participant behavior and are fit using the \texttt{glmmTMB} package (AGPL-3 License) in R \cite{brooks-glmmtmb, mcgilly-glmmtmb}. 

To analyze the likelihood of question-asking, we model the presence of a question mark in a contribution as a binary outcome following a Bernoulli distribution

$$ Y_{ij} \sim \text{Bernoulli}(p_{ij})$$
$$ \text{logit}(p_{ij}) = \beta_0 + \beta_{1}\text{T1}_j + \beta_{2}\text{T2}_j + u_j $$
where $Y_{ij}$ represents the binary outcome for contribution $i$ from participant $j$, $p_{ij}$ is the probability of at least one question mark in the contribution, and $\beta_{1}$ and $\beta_{2}$ are the fixed effects for the treatment arms. The term $u_j$ is the random intercept for participant $j$, where $u_j \sim N(0, \sigma^2_u)$ with between-participant variance $\sigma_u^2$ capturing individual heterogeneity in question mark usage. 

The second text-based metric we analyzed to assess this research question is the Perspective API curiosity score. The Perspective API is a machine-learning-based tool developed by Jigsaw and Google to analyze the attributes of digital conversation \cite{lees-perspective}.  Its curiosity score specifically measures the degree to which a post ``attempts to clarify or ask follow-up questions to better understand another person or idea.'' Scores range between 0 and 1, indicating the \textit{probability} that a reader will find the text to exhibit this property. Using \texttt{v1} of the \texttt{Perspective Comment Analyzer API} in Python, we measured curiosity scores of all user contributions.

To analyze the curiosity scores, which range from 0 to 1, we used a Beta Mixed-Effects Model. While scores are often used for thresholding, direct modeling of the scores also proves useful for comparing texts \cite{colglazier_effects_2024}. 
Since the Perspective API is sensitive to punctuation, and punctuation has been shown to be able to drastically alter scores \cite{hosseini_deceiving_2017}, we included the number of question marks ($Q$) as a covariate to ensure we are not simply modeling any change in question mark usage. The beta regression with a logit link assumes the curiosity score $C_{ij}$ for contribution $i$ from participant $j$ follows a Beta distribution defined by the mean $\mu_{ij}$ and a dispersion parameter $\phi$
$$ C_{ij} \sim \text{Beta}(\mu_{ij}, \phi) $$
$$ \text{logit}(\mu_{ij}) = \beta_0 + \beta_{1}\text{T1}_j + \beta_{2}\text{T2}_j + \beta_{3}{Q}_{ij} + u_j $$
As before, $u_j \sim N(0, \sigma^2_u)$ is a participant random intercept; here,  between-participant variance $\sigma_u^2$ captures individual heterogeneity in curiosity. The regression coefficients $\beta_1$ and $\beta_2$ correspond to the additive change in the log-odds of the expected score for each treatment group relative to the control. We exponentiate coefficients to find the multiplicative change in the odds ratio of the expected score, $\frac{\mu_{ij}}{1-\mu_{ij}}$.

\textit{\textbf{RQ2: Toxicity.}} Does curiosity priming reduce toxicity in user contributions? Beyond increasing curiosity, a goal of our intervention was to determine if curiosity priming could mitigate antisocial behavior on a platform. To evaluate this, we examined the toxicity of user posts as measured by the Perspective API. The toxicity score represents the probability that a reader would perceive the post as toxic, defined by the API as ``a rude, disrespectful, or unreasonable comment that is likely to make you leave a discussion''. 

To analyze these scores, we used the same Beta Mixed-Effects modeling framework described in RQ1. This approach is again chosen to handle the bounded, non-normal distribution of toxicity probabilities while accounting for participant-level random intercepts.

\textit{\textbf{RQ3: Volume of engagement.}}  Does curiosity priming affect overall engagement volume? To evaluate the overall impact of curiosity priming on user participation, we analyzed the volume of interactive activity across the three treatment arms. Specifically, we measured the number of comments, the cumulative time spent composing comments, and the number of engagements by each user in each treatment arm. To evaluate statistical differences across the three conditions, we conducted one-way ANOVA tests. We used Tukey-HSD tests as a post-hoc analysis to determine which groups' differences in means drive any significant effects. This approach makes all pairwise comparisons while controlling for the increased chances of Type I errors with multiple testing. 

\textit{\textbf{RQ4: App perception.}} Does treatment affect how much users enjoy the app? To evaluate whether the introduction of curiosity-priming features impacted the user experience, we measured participants' perceptions of the platform through a post-treatment survey. Specifically, users were asked to rate the degree to which the word ``enjoyable'' described their experience with the application. Responses were collected on a four-point scale: ``extremely well,'' ``very well,'' ``somewhat well,'' and ``not at all well.'' To determine if the treatments influenced user enjoyment, we used Chi-square tests of independence to compare the distribution of categorical ratings between the treatment groups and the control group.

\textit{\textbf{RQ5: Intellectual humility (IH).}} Does interacting on a curiosity-primed treatment arm change participants' measured intellectual humility? To measure how treatment impacts participants' attitudes, we measured their intellectual humility in pre- and post-treatment surveys using the six point scale \cite{leary_cognitive_2017}. They rated how well the six phrases below applied to themselves on a five point Likerts scale from ``not well at all'' (1) to ``extremely well'' (5):

\begin{itemize}
    \item I question my own opinions, positions, and viewpoints because they could be wrong.
    \item I reconsider my opinions when presented with new evidence.
    \item I recognize the value in opinions that are different from my own.
    \item I accept that my beliefs and attitudes may be wrong.
    \item In the face of conflicting evidence, I am open to changing my opinions.
    \item I like finding out new information that differs from what I already think is true.
\end{itemize}

Adding their responses from each question, each participant received a total intellectual humility score from 5 to 30. We then computed within-subject change in intellectual humility (pre-treatment minus post-treatment). We conducted a one-way ANOVA test to determine if there is any difference in the change in intellectual humility between treatment arms. 

Collectively, the outcomes from the five research questions allow us to assess whether curiosity-priming interventions can successfully shift user attention and improve discourse quality without compromising the overall user experience.

\section{Results}

Across the three arms there were 8,535 comments and 2,470 posts written by human participants, alongside 90,661 comments and 35,657 posts generated by our AI bots. The high ratio of bot to human content underscores that participants primarily consumed rather than produced content. Table~\ref{tab:results} summarizes key descriptive statistics relevant to each research question by treatment arm.

\begin{table}[h]
\centering
\footnotesize 
\renewcommand{\arraystretch}{1.1} 
\begin{tabular}{l @{\hspace{0.1em}} ccc}
\toprule
\textbf{Metric} & \textbf{Control} & \textbf{T1} & \textbf{T2} \\
\midrule
\textbf{Sample Sizes} \\
Participants & 773 & 740 & 769 \\
Posts / Comments & 784 / 3.2k & 826 / 2.6k & 860 / 2.7k \\
\midrule
\textbf{RQ1: Curiosity} \\
\% with $\geq$ one ``?'' & & &  \\
\quad *Posts & 52 & 74.7  & 79 \\
\quad Comments & 9.8 & 10.5  & 10.3 \\
Mean (SD) Curiosity & & &  \\ 
\quad *Posts & 0.60 (.32) & 0.77 (.23) & 0.80 (.19) \\
\quad Comments & 0.41 (.25) & 0.43 (.25) & 0.42 (.25) \\
\midrule
\textbf{RQ2: Toxicity} \\
Mean (SD) Toxicity & & &  \\ 
\quad *Posts & .043 (.08) & .028 (.05) & .03 (.05) \\
\quad Comments & .04 (.07) & .036 (.06) & .039 (.07) \\
\midrule
\textbf{RQ3: Volume} \\
Mean (SD) \# / User & & &  \\ 
\quad Posts & 1.38 (.73) & 1.34 (.72) & 1.36 (.70) \\
\quad *Comments & 4.64 (3.13) & 4.11 (2.7) & 4.08 (2.99) \\
\quad *Engagement & 4.56 (3.51) & 4.38 (3.51) & 3.98 (2.93) \\
Mean (SD) Time & & &  \\ 
\quad Posts & 88.4 (84.5) & 91.1 (83.8) & 88.8 (82.2) \\
\quad Comments & 208 (133) & 194 (129) & 199 (136) \\
\midrule
\textbf{RQ4: Enjoyment } \\
Mode Ans. (\%) & Very (37.6) & Very (38.6) & Very (37.2) \\ 
\midrule
\textbf{RQ5: IH } \\
Mean (SD) IH Diff. & -.36 (2.7) &-0.5 (2.51) & -.46 (2.69) \\

\bottomrule
\end{tabular}
\caption{Descriptive statistics by treatment arm. Asterisks (*) denote features with significant treatment effects at a 0.05 level using the methods described prior.}
\label{tab:results}
\end{table}

\subsection{RQ1: Curiosity}

We first compare posts and comments across conditions looking for behavioral evidence of curiosity. In the control arm, 52\% of user posts contain at least one question mark. With curiosity norms (T1), 74.7\% of user posts contained question marks; with norms and affordances (T2), this was 79\% The high baseline level of question marks in the control arm no doubt reflects that bot question asking encourages environment of question asking. Still, despite the high baseline, we see increases in the frequency of question mark usage in two treatment arms. 

\begin{figure}
    \centering
    \includegraphics[width=\linewidth]{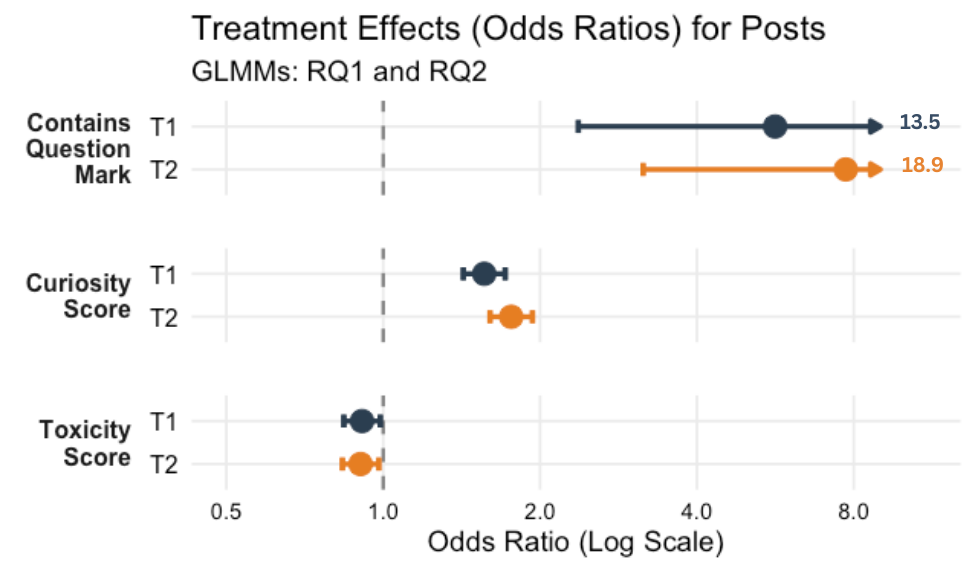}
    \caption{Points represent Odds Ratios ($e^{\beta}$) from treatment coefficients for T1 (blue) and T2 (orange) in GLMMs. Error bars indicate 95\% confidence intervals; arrows signify that the confidence interval extends beyond the plotted axis range. Contains Question Mark reports effects from logistic GLMM; Curiosity and Toxicity Scores report effects from Beta GLMMs. An Odds Ratio greater (less) than 1 represents an increase (decrease) relative to the control group. }
    \label{fig:betas}
\end{figure}

Our GLMM analysis confirms these behavioral changes are statistically significant. 
Compared to the control group, participants in both treatment arms exhibited a significantly higher likelihood of including a question mark in their posts (T1: $\beta_1$ = 1.73, $p = 1 \times 10^{-4}$; T2: $\beta_2=2.04, p = 8.3 \times 10^{-6}$). These coefficients correspond to a substantial increase in the odds of question-asking. Controlling for participant-level variation in question mark usage, participants in T1 and T2 have 5.6 and 7.7 times higher odds of using a question mark, respectively, than those in the control group. Figure \ref{fig:betas} shows these point estimates as well as 95\% confidence intervals.

Using the Perspective API, we see significantly higher curiosity scores from each treatment arm as compared to the control group (T1: $\beta_1$ = 0.45, $p < 10^{-16}$; T2: $\beta_2= 0.57, p < 10^{-16}$). Adjusting for individual level variation and question mark usage, T1 and T2 corresponded to a 1.57 and 1.77 times increase in the odds of the expected curiosity score, as is shown in Figure \ref{fig:betas}. Together, the question mark and curiosity score treatment effects provide evidence that users' posts exhibited higher measures of curiosity when primed with curiosity norms or norms and affordances. 

Similar results are not found in comments: in all treatment arms, close to 10\% have question marks. We fit the same models used for posts to user comments and found that question mark usage and curiosity score are similar across all arms with no significant differences.

\subsection{RQ2: Toxicity} 

To evaluate antisocial behavior, we analyzed Perspective API toxicity scores. We note that the average toxicity scores for posts and comments across all treatment arms were low, as is shown in Table \ref{tab:results}. Nevertheless, using the Beta Mixed-Effects, we find significant effects from treatment arm on these scores. Treatment had the opposite effect on toxicity from curiosity: participants in T1 and T2 showed lower likelihoods for post toxicity ($ \beta_1 = -0.094, p = 0.023$; $ \beta_2 = -0.10, p = 0.015$) compared to the control arm. Figure \ref{fig:betas} shows that adjusting for individual level variation and question mark usage, T1 and T2 corresponded to a 0.9 and 0.87 times decrease in the odds of the expected curiosity score. For both curiosity and toxicity, we see that the point estimates for T2 correspond to slightly stronger effects than T1. However, their overlapping confidence intervals imply that these differences are not significant. For both T1 and T2, the treatment effects on the curiosity score are stronger (further from 1) than on the toxicity score. 
These results suggest that user posts primed with curiosity norms or norms and affordances exhibit lower toxicity, a measure of antisocial behavior.
As with curiosity, fitting the same models to user comments yielded no significant differences in toxicity across arms. 

\subsection{RQ3: Engagement Volume}

Table 2 highlights a decrease in interaction frequency for users in the treatment arms. Specifically, while control users averaged 4.64 comments and 4.56 engagements, these dropped to 4.11 comments and 4.38 engagements in T1 and 4.08 comments and 3.98 engagements in T2. One-way ANOVA tests confirmed there were statistically significant differences for both comments per user ($p =  2.6 \times 10 ^ {-5}$) and engagements per user ($p = 0.01$). Post-hoc Tukey-HSD analysis shows that both T1 and T2 have a significantly different number of comments than the control arm ($p < 0.0001$ for each), but are not significantly different from each other. Only T2 had a significantly different number of engagements from the control arm ($p = 0.006$). 

No significant differences were found in the total number of original posts per user or the total cumulative time spent writing comments or posts per user. This suggests that while curiosity priming might reduce the quantity of interactive behavior, it does not diminish original contributions or the overall temporal effort users invest in the platform.

\subsection{RQ4: App Perception}

We found no significant difference in the perception of the app between users in the treatment groups and those in the control group. A majority of the participants in each treatment arm found app at least somewhat enjoyable, with over half rating extremely or very. Chi-square tests of independence revealed no significant differences in the distribution of ratings between either of the treatment groups and the control group. 
This null result is promising: it suggests that introducing curiosity priming features, even those that alter the interface design, did not render the experience less enjoyable than the more standard social media environments users are more accustomed to.

\subsection{RQ5: Intellectual Humility (IH)}

The median difference in intellectual humility for users in all three treatment arms was zero; Table \ref{tab:results} shows that the mean difference in each treatment arm are slightly negative. Using a one-way ANOVA test, we found no significant difference in users' change in intellectual humility between any treatment arms.

\section{Discussion and Conclusion}

Efforts to promote prosocial behavior on social media face a fundamental challenge: researchers typically cannot manipulate the platform features that shape user behavior, making causal inference difficult. Our study addresses this challenge by conducting a randomized controlled trial on a new research platform augmented with AI bots to emulate realistic social media dynamics. This approach enabled us to systematically test whether curiosity-priming interventions implemented through social normative cues and interface affordances could shift user behavior toward more inquisitive and less toxic discourse. Our findings demonstrate that such interventions can successfully encourage prosocial behaviors, offering both theoretical insights into curiosity as a mechanism for improving online discourse and practical implications for platform design.

The most striking behavioral change occurred in our measures of question-asking and curiosity. Participants exposed to curiosity priming through normative cues alone (T1) or combined with interface affordances (T2) asked significantly more questions in their original posts compared to control participants. Perspective API curiosity scores similarly increased across both treatment arms. These results provide evidence that curiosity can be primed through relatively lightweight platform interventions, shifting users from declarative or argumentative modes of engagement toward more exploratory and receptive communication. For consistency across experimental conditions, the bots in our study write posts that are phrased as questions in all conditions. As a result, the base rate of question-asking displayed by our respondents is already high, suggesting our treatment effects could be conservative. The observed treatment effects were limited to original posts rather than comments. This pattern may reflect the nature of commenting as inherently responsive, in that users answer questions rather than posing new ones, though it could also indicate that 15-minute sessions provided insufficient time for curiosity norms to permeate deeper into conversational behavior. The question of whether cued norms such as curiosity diffuse through social interactions over time remains an important avenue for future research.

Our interventions also reduced toxicity in user posts. While baseline toxicity was low in our controlled environment, partly due to the absence of intentionally inflammatory content, the significant decrease across both treatment arms suggests that curiosity priming mitigates this antisocial behavior. This finding aligns with theoretical accounts linking epistemic curiosity to intellectual humility and reduced defensiveness \cite{leary_cognitive_2017}. When users approach online discourse with the goal of learning rather than winning arguments, perhaps hostile or contemptuous language becomes less functional. The observed negative correlation between question marks and toxicity scores further supports the interpretation that posts formulated as questions are structurally less conducive to toxicity. Whether these effects would persist in online spaces with higher baseline toxicity is an empirical question that future work should address, particularly as commercial platforms can often feature more aggressive and politically-charged discourse than our experiment.

We would emphasize that this experiment was conducted in an especially polarized and emotionally charged political context, just before the 2024 U.S. Presidential election. We might expect interventions promoting curiosity around political topics would face a challenge in such an environment. The fact that our primes still produced significant effects on question-asking and toxicity suggests promise in even a lightweight platform design change aimed at promoting curiosity. 

The observed reduction in commenting and liking behaviors warrants attention. Despite a decrease in the volume of liking and commenting actions, we found no overall decrease in time spent writing posts and comments. This suggests that users may have shifted from high-frequency, low-effort actions toward fewer but more thoughtful contributions. However, our current measures cannot fully adjudicate this interpretation. Dwell time, which would indicate whether users spent more time reading and processing content even when they interacted less frequently, remains difficult to measure and interpret in multi-post feeds where attention is distributed across visible content \cite{epstein2022quantifying}. Future research incorporating more granular attention metrics, perhaps through integrated recall checks or eye-tracking, could clarify whether reduced engagement volume reflects deeper cognitive processing or broad disengagement.

Although our behavioral measures of curiosity were responsive to the interventions, we found no attitudinal differences across conditions as measured in the post treatment survey. The null effect on intellectual humility is theoretically informative, suggesting that momentary boosts in curious behavior do not immediately result in dispositional changes of epistemic attitudes. This dissociation between state-level behaviors and trait-level dispositions is consistent with social-cognitive theories distinguishing situational influences from stable individual differences \cite{bandura_social_1977}. In practical terms, it indicates that fostering curiosity-driven interactions may require sustained treatment rather than brief priming. At the same time, the fact that participants found the curiosity-primed environment equally enjoyable as the control is encouraging. It suggests that interventions promoting prosocial discourse do not necessarily detract from the user experience---a concern often raised when proposing design changes that might reduce engagement metrics platforms use to measure success.

While our findings suggest the potential for social media to be redesigned for social good, our study does have limitations that warrant discussion. 

 First, our study was conducted in a controlled 15-minute session with participants recruited for research purposes, not in the midst of organic social media use. Whether curiosity norms would persist under the pressures of real-world platforms, with their algorithmic feeds, social network effects, and more emotionally valent content, remains uncertain. The bot prompts ensured that participants experienced a largely non-toxic conversational environment and did not face the full range of antagonistic or emotionally charged interactions characteristic of open platforms that could impact how they would respond to the treatment. Similarly, our topic domain of energy and climate was chosen to be substantive but not maximally polarizing. Effects might differ for more contentious topics like abortion or immigration. Finally, we measured immediate behavioral outcomes but not longer-term consequences such as belief updating, network diversity, or sustained participation. Whether curiosity-primed interactions lead to deeper epistemic gains or simply more polite disagreement is an open question.

\subsection{Implications for Platform Design}
Beyond the substantive implications of our findings, this study contributes to an emerging research agenda focused on prosocial platform design. While much attention has been paid to how social media amplifies affective polarization, outrage, and echo chambers, less work has systematically tested interventions to counteract these dynamics. We offer both an approach to testing design ideas and an initial look at curiosity as a mechanism for increasing prosocial behavior on social media. We evince that platforms can be intentionally designed to promote prosocial behaviors without sacrificing user experience. By leveraging experimental control made possible through independent research platforms, we provide causal evidence that curiosity priming increases question-asking, reduces toxicity, and shifts engagement patterns toward more deliberate interaction online.  Unlike interventions requiring extensive content moderation effort or algorithm redesign, the curiosity intervention studied here can be implemented through relatively simple changes to onboarding language and interface elements. These findings offer a foundation for future work exploring how digital environments can be structured to support healthier, more productive public discourse. As researchers continue to grapple with the challenges of studying social behavior on proprietary platforms, methods that balance experimental control with realistic social dynamics will be essential for translating findings into actionable platform interventions.

\subsection{Ethical Statement}
Our study was approved by the IRB\footnote{institution and protocol number omitted for peer review}, and we acquired explicit consent from all participants (see Appendix~\ref{app:consent}). In the consent, participants were informed that they might interact with bots and the content they were exposed to is similar to regular social media platforms (see Appendix~\ref{app:bots}). The discussed topics were of no particular concern for the U.S. election to not risk affecting voter opinions. We recruited the anonymous participants on Prolific (see Appendix~\ref{app:recruitment}), who are only identifiable by ID. While we collected basic demographic data like age and gender, we did not collect personally identifiable information. The recruitment, survey, and platform data are stored in approved systems and only linked when necessary for analysis. We only report aggregate results and refrain from releasing any textual content generated by participants to avoid any potential deanonymization. Therefore, the potential harm to human subjects is negligible. 

We used a commercial large language model from OpenAI, which is subject to inherent biases and might produce hallucinations.  The risk of these biases affecting the behavior is limited due to the short study duration. Similarly, the risk of model shift (i.e., the tendency of these models to behave differently over time) affecting the outcome is again confined by all participants taking the study in a narrow period (less than 24 hours) and being exposed to the same control conditions as a starting point (i.e., the same feed and AI bots). 

Our research focused on prosocial behavior in social media, which has the potential for societal benefits. Noteworthy, our results indicate that the interventions had no detrimental effect on app perception while increasing curiosity and decreasing toxicity. Still, we observe that curiosity primes reduce quick engagements such as liking and commenting to some degree. Without anticipating such reductions in the design of recommender systems, a platform may inadvertently de-prioritize curious, good-faith content as a result of these signals. Overall, we see the epistemic benefits of our study as a net positive, as our study shows a path forward for more productive social media environments.

\bibliography{curiosity_paper}
\appendix

\section{Synthetic Profiles and Content Feed}
\label{app:bots}

All participants are subject to the same initial social media environment, including synthetic profiles and content feed. Each synthetic profile comprises a name and a persona. The persona includes the occupation and three distinct dimensions (2 on each dimension): 
\begin{itemize}
    \item Political: Republican, Democrat, Independent
    \item Climate Change Knowledge behavior: overconfident, reserved, well-informed
    \item Climate Corrective Actions attitude: 
pessimistic (against actions), realistic (for impactful actions), optimistic (pro actions)
\end{itemize}
Table \ref{tab:my_label} provides the full characteristics for all simulated bot profiles, including engagements on the prefilled and scheduled content feed. 
\begin{table*}[htb]
    \centering
    \begin{tabular}{l|l|l|l}
    profile ID - profile name & persona: occupation (political, knowledge, attitude) & engagements$\dagger$ \\ \hline
P1 - RealDataFlow & 		Climate Scientist (Democrat, Well-informed, Realistic) 		& 0p 4c 2l 0d\\
P2 - GreenMindset33 & 	Vegan Student (Democrat, Overconfident, Optimistic) 		& 2p 1c 1l 0d\\
P3 - Ledger\_Logic &		Tax Accountant (Republican, Reserved, Pessimistic) 		& 2p 0c 0l 0d\\
P4 - HustleAndHeft! & 		Fitness Coach (Republican, Overconfident, Optimistic) 		& 1p 1c 1l 1d\\
P5 - Artisanview &		Master Craftsman (Independent, Reserved, Realistic) 		& 2p 0c 0l 0d\\
P6 - C0deAndC0nsider & 	Software Developer (Independent, Well-informed, Pessimistic) 	& 1p 2c 1l 2d\\
    \end{tabular}
    \caption{$\dagger$ assumed engagements (p: \# of posts, c: \# of comments, l: \# of likes, d: \# of dislikes)}
    \label{tab:my_label}
\end{table*}

 	
	
	


The scheduled and prefilled feed all users see when they enter the app was in total: (8 posts, 8 comments, 5 likes, 3 dislikes).  
Below we present the feed (in reverse chronological order) using the following format:
\begin{itemize}
    \item $\text{T}\pm x$ with $x$ describing the relative time in the feed in minutes.
    \item $\text{-c} n$ optional if the post has comments and $n$ specifying the position of the comment.
    \item $@profile$ specifies the author's name of the synthetic profile.
    \item the textual content
    \item $<$engagements$>$ optionally if there are engagements on the post, describing the assumed engagements by the synthetic profiles grouped by type
\end{itemize}

\noindent
Scheduled feed:
\begin{quote}
T+5: @Ledger\_Logic What really bothers me is that fossil fuel companies get all these subsidies with my taxpayer money. Wouldn’t ending these be an easy-to-implement solution to a major part of the problem and benefit us all?\\
T+3: @GreenMindset Looking at the election, is there even a choice when you support climate action?\\
T+2: @HustleAndHeft! Does switching to a vegan diet really help? I really like meat but wondering if it would make a difference.\\
\end{quote}

\noindent
Prefilled feed:
\begin{quote}
T-1: @Artisanview Do we need multiple solutions to face climate change? Wouldn’t it be enough to just switch to green energy and electrical for everything?\\
T-5: @Ledger\_Logic I try to reduce my impact by conserving energy wherever possible (having the lid on the pot, turning off the lights), as well as recycling my cardboard boxes, cans, etc. but it seems like so little impact. What else can I personally do to reduce my carbon footprint?	$<$disliked by P6$>$\\
T-5-c1: @RealDataFlow This is a good start, but cannot replace more impactful footprint reductions, like in the area of transportation.\\
T-5-c2: @C0deAndC0nsider Indeed, the carbon footprint of switching off the light (assuming you have energy-saving bulbs) is negligible. Also, how well recycling works is debatable. Turns out a lot of it gets thrown away.\\
T-7: @C0deAndC0nsider What would be a good strategy to reduce my negative impact on the environment?	<liked by P2>\\
T-7-c1: @HustleAndHeft! If possible, switch to cycling. For instance, on a mild day, it might be pleasant to cycle to work, and it will keep you fit at the same time. \\
T-10: @GreenMindset33 What is an activity with a high carbon footprint?	$<$liked by P1,P3,P5 disliked by P4, P6$>$\\
T-10-c1: @RealDataFlow It depends on what you mean by high. While you might have a higher overall footprint by driving your car every day, flying just once already emits a lot of CO2.\\
T-10-c2: @GreenMindset33 So, which of the two would you focus on reducing?\\
T-10-c3: @RealDataFlow Both if possible.\\
T-10-c4: @C0deAndC0nsider Just get rid of dirty industry in the first place, no other sector produces such a high amount of CO2.\\
T-10-c4: @RealDataFlow Actually, the two highest sectors are transportation and electricity production. Industry is third.\\
T-11: @Artisanview Are people taking advantage of the Biden tax credits for  n simple solutions like insulating homes and switching your old boilers for heat pumps?		$<$liked by P4$>$\\
T-12: @GreenMindset33 Why is there so little research on building greener communities? Concepts like vertical farming seem promising. 		$<$liked by P1, P6$>$\\
T-12-c1: @C0deAndC0nsider I would not say that there is little research, just that it is not recognized by politicians. They don’t seem to give a f***... I wish they were talking more about climate change this election.\\
T-12-c2: @RealDataFlow It is indeed an unfortunate problem overall. That there is so little action, in many areas, although the research is very clear.\\
\end{quote}

Besides these initial (scheduled) posts, comments, and engagements, the synthetic users are configured to also dynamically generate new posts, comments, and engagements that adhere to their specified persona.

\section{Participant Recruitment}
\label{app:recruitment}

We used Prolific for recruitment. We aimed for a reward of \$7.5 for of 30min study, with the potential to receive a \$1 bonus for answering a question in the posttreatment correctly. 

We ran a pilot study on 30th of October 2024 that took 4.5 hours total, resulting in 294 approved submissions (with an additional 5 rejected; 47 returned; 4 timed-out). The median time to complete the pilot was 28 minutes, with an average hourly reward of 
\$16.12 (including 66 bonus payments). The pilot was performed to test the study design and technical feasibility. The total cost of the pilot, including platform fees, was \$3,028.00. 

The actual study was run on the 3rd of November 2024, which took approximately a day (around 23 hours in total) to receive 2218 submissions (with an additional 23 rejected; 3580 returned; 40 timed-out). The median time complete was again 28 minutes with an average hourly reward of \$15.99 (including 532 bonus payments). The total cost of the study, including platform fees, was \$22,889.34.

\begin{figure*}[t]
    \centering
    \includegraphics[width=\linewidth]{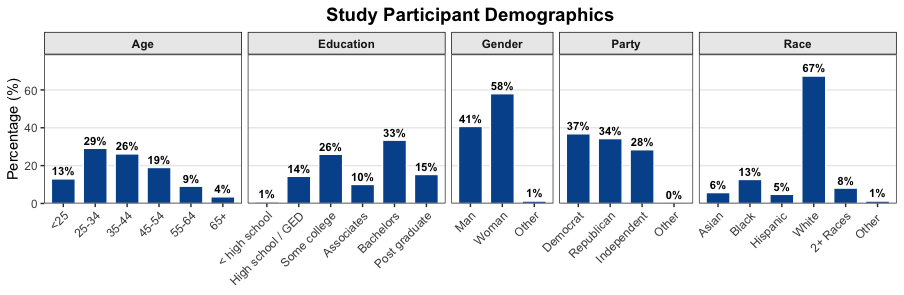}
    \caption{Demographic Distributions of the Participants.}
    \label{fig:demographics}
\end{figure*}

Figure \ref{fig:demographics} provides the breakdown of participant demographics.
 
Recruitment message with the title ``Help test a new social media platform!'': 
\begin{quote}
    We would like to know your opinions about a new social media platform. Spend 30 minutes trying out the platform and sharing your thoughts!
    
    Click the link below to be taken to an intake survey. Once complete, you will be given instructions on how to get access to the platform through the “Spark Social” testing environment. You will need to download “Spark Social” onto your mobile device to complete this task.
    
    To be eligible for this study, you must:
    1. Be a US resident.
    2. Be at least 18 years of age.
    3. Own an Apple iPhone or an Android device.
    4. Be willing to download “Spark Social” onto your smart phone.
    To ensure that you are compensated, please leave this window open as you complete this task!
\end{quote}

\section{Consent Form}
\label{app:consent}

\begin{quote}
Thank you for your interest in this research by $<$blinded$>$ aimed at improving social media experiences and features. We want to learn your views of a new mobile app, which we are evaluating on a social media testing environment called Spark Social. 

To participate, you will first be asked to complete a 5-minute survey to ensure you meet the eligibility criteria for our study.  You will then be given instructions to download Spark Social from the Apple or Google Store.  After using the platform for 15 minutes, we will ask some questions about your experiences and opinions. Testing the platform and providing your feedback should, in total, take no longer than 30 minutes of your time. You will receive \$7.50 for completing the study. To encourage careful engagement on the platform, a bonus of \$1 is available for correctly answering a question at the end. 

Your participation is completely voluntary, and you may withdraw at any time. However, you must meet each of the following criteria to receive a full payment of \$7.50:
 
You must be at least 18 years of age.
You must reside in the U.S.
You must own an Apple iPhone or an Android device (a smart phone).
You must download the Spark Social testing environment.
You must complete questions about your views and experience on the platform and successfully pass attention checks.

While all research faces some risk of loss of confidentiality, we have designed this research to minimize that risk and to always maintain your confidentiality. We never ask for your name or any other information that might identify you. The Spark Social app will collect information as you navigate it but will NOT collect any personally identifiable information (or record your voice or video). Although collected data may be made public or used in the future, your identity will always remain confidential.

You may encounter a range of different views on our platform, which is currently hosted on the Spark Social testing environment. Just like on any other social media platform, some of the content you will see will have been generated by artificial intelligence algorithms (bots). It’s also possible that you may disagree with the opinions of others or the way they are expressed. Please know that you have the option to block any users you wish, at any time.

For answers to any questions you may have about our platform or research, please contact $<$blinded$>$ at $<$blinded$>$. For questions about your rights as a participant, contact the $<$blinded$>$ Institutional Review Board at $<$blinded$>$. Please reference Protocol ID $<$blinded$>$ in your email. 

I am a US resident, at least 18 years of age, and desire of my own free will to participate in this research.
\end{quote}

\section{Instructions and Conditions}
\label{app:arms}

\begin{figure}[t!]
    \centering
    \begin{subfigure}[t]{0.5\textwidth}
        \centering
        \includegraphics[width=0.99\textwidth]{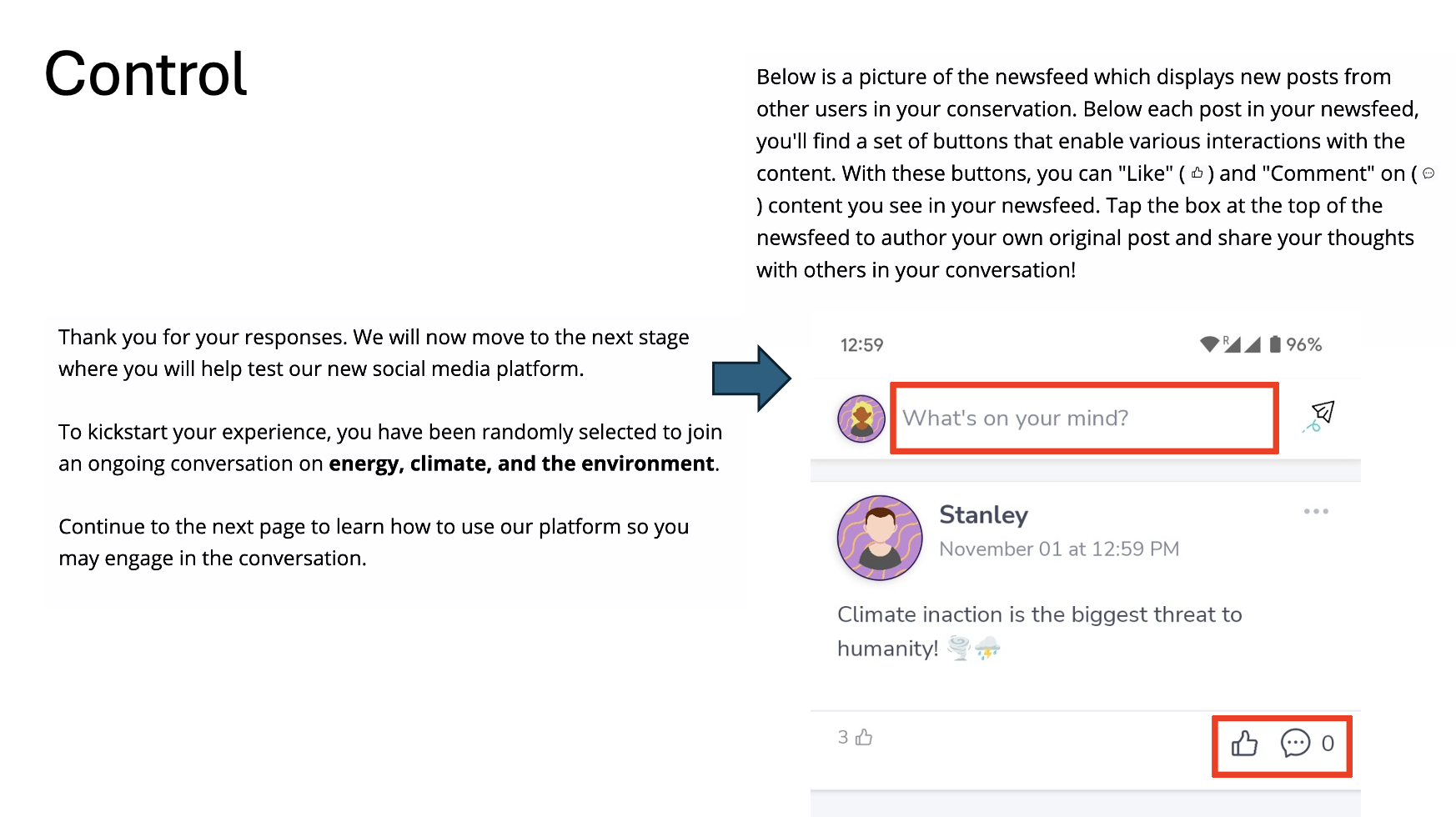}
        \caption{Control Condition - No Curiosity Priming.}
    \end{subfigure}\\
    \begin{subfigure}[t]{0.5\textwidth}
        \centering
        \includegraphics[width=0.99\textwidth]{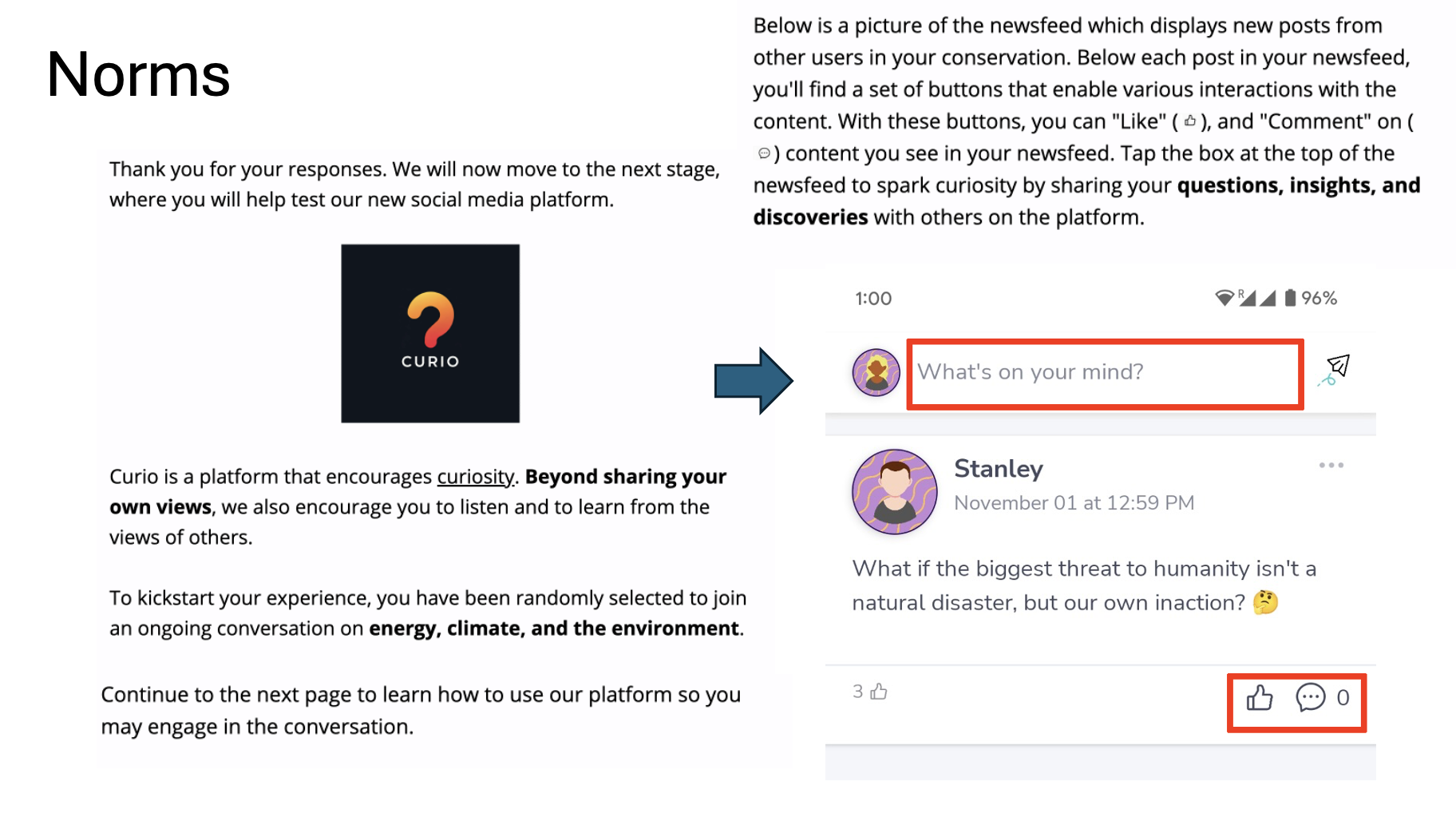}
        \caption{Treatment T1: Curiosity Norms and Affordances.}
    \end{subfigure}\\
    \begin{subfigure}[t]{0.5\textwidth}
        \centering
        \includegraphics[width=0.99\textwidth]{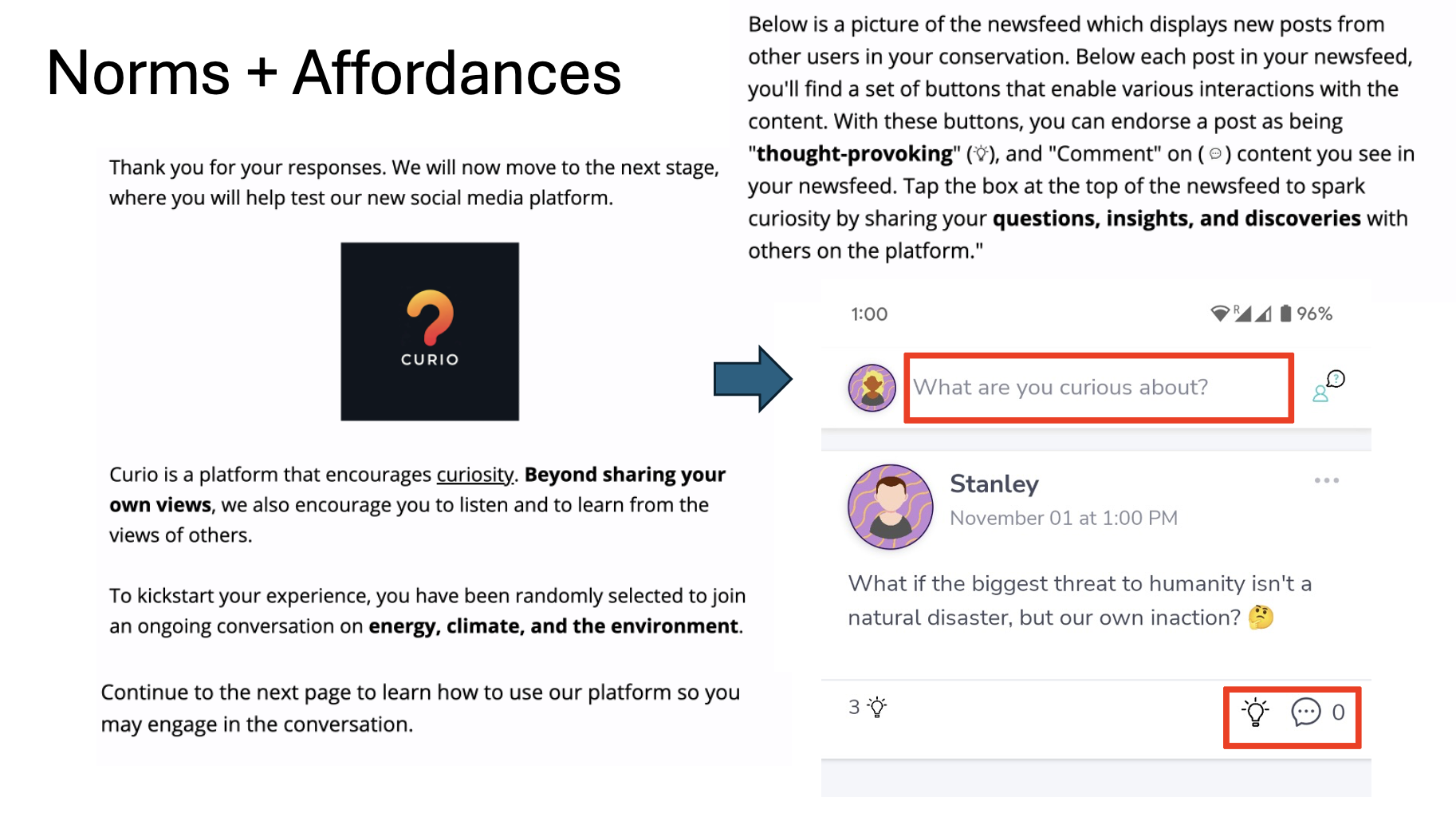}
        \caption{Treatment T2: Curiosity Norms and Affordances.}
    \end{subfigure}
    \caption{We prime participants with two-page instructions before they enter the newsfeed.}
    \label{fig:setup}
\end{figure}

Figure \ref{fig:setup} provides the two pages of instructions provided to participants in each of the three treatment arms. In all three arms, participants are informed on the first page that they will be testing a social media platform and discussing climate, energy, and the environment. In the two treatment arms, they are provided with an app logo shaped like a question mark and the app name ``Curio". No specific app name or features are provided in the control arm. Additionally, the treatment arms specifically inform users that curiosity is encouraged.

The second page provides more detailed instructions on how to use the app. In the control arm and T1 (norms) treatment arm, users are prompted to enter text with the phrase ``What's on your mind`` next to a small paper airplane button. In the T2 (norms and affordances) arm, the text is changed to ``What are you curious about?" and the icon is replaced with a question mark in a thought bubble. Additionally, this arm replaced the standard like button with a light-bulb icon.

\section{Preregistration}
\label{app:prereg}

We preregistered the study design on OSF.\footnote{due to an expired embargo period, we cannot share the link anonymously and will add it after acceptance} We summarize the relevant points for the study below. The recruitment procedure described in the preregistration (see Appendix \ref{app:recruitment}) was implemented as described, but we did have some minor deviations in the analysis, including the use of GLMM models. Given the reporting of some exploratory analyses (as preregistered), we report two-sided tests throughout.

Preregistered Hypotheses:
\begin{itemize}
    \item Participants assigned to a curiosity treatment will generate more comments and engagements. \emph{We observed the opposite tendency.}
    \item Participants assigned to a curiosity treatment produce less toxic posts and comments. \emph{We only observed less toxicity on the posts.}
    \item Participants assigned to one of the studies two treatment conditions will perceive the app more positively, i.e., associate more positive and fewer negative words with the platform. \emph{We did not find a significant difference.} 
    \item Participants assigned to a curiosity treatment will report higher intellectual humility. \emph{We did not observe any significant changes.}
\end{itemize}

\noindent
Intake Survey Items:
\begin{itemize}
    \item \textbf{Age} ‘What is your age in years?’ [Open-ended, integers between 0 and 120 allowed]
    \item \textbf{Education} 'What is the highest level of school that you have completed or the highest degree you have received?' [6-point scale from 'Less than high school’ to ‘Post-graduate degree’]
    \item \textbf{Race/Ethnicity} ‘What racial/ethnic group(s) do you consider yourself to be a part of? Select all that apply’ [Asian; Black or African American; Hispanic; White; Something else]
    \item \textbf{Gender} ‘Which of the following best describes your gender?’ [Woman; Man; Something else; Prefer not to say]
    \item \textbf{PID} ‘Do you usually think of yourself as a Democrat, a Republican, an Independent, or what?”
\end{itemize}

\noindent
Pretreatment Survey Items:
\begin{itemize}
    \item \textbf{IH} ‘How well, if at all, do the following phrases apply to you?’ [I question my own opinions, positions, and viewpoints because they could be wrong/I reconsider my opinions when presented with new evidence/I recognize the value in opinions that are different from my own/I accept that my beliefs and attitudes may be wrong/In the face of conflicting evidence, I am open to changing my opinions/I like finding out new information that differs from what I already think is true.] [5-point scale from 'Not at all well' to 'Extremely well']
\end{itemize}

\noindent
Posttreatment Survey Items:
\begin{itemize}
    \item \textbf{Platform Experience} Below are several words that you may or may not use to describe your experience using our new social media platform. For each word, please indicate the degree to which the word does or does not describe your experience on the platform. [4-point scale of Not well at all, Somewhat well, Very well, Extremely well] [Word list: Engaging, Entertaining, Toxic, Enjoyable, Frustrating]
    \item \textbf{IH} ‘How well, if at all, do the following phrases apply to you?’ [I question my own opinions, positions, and viewpoints because they could be wrong/I reconsider my opinions when presented with new evidence/I recognize the value in opinions that are different from my own/I accept that my beliefs and attitudes may be wrong/In the face of conflicting evidence, I am open to changing my opinions/I like finding out new information that differs from what I already think is true.] [5-point scale from 'Not at all well' to 'Extremely well']
\end{itemize}







\section{Data}
\label{app:data}

We emphasize that our study design requires different components (i.e., recruitment platform, survey platform, our independent research platform) integrate to ensure anonymity of participants (e.g., no identifiable Prolific ID will be sent to the research platform) and will only be linked for analysis when necessary. For that reason, there can be different numbers of records, e.g., when a participant completes a study on Prolific before answering all the posttreatment survey items. 

Data is uploaded alongside the submission; upon publication, a public repository containing the data will be made available.

\end{document}